\documentclass[aoas]{imsart}

\usepackage{amsmath, amsfonts, amssymb}
\usepackage[authoryear]{natbib}
\usepackage{graphicx}
\usepackage{xcolor}
\usepackage{gensymb}
\usepackage{subcaption}
\usepackage[version=4]{mhchem}
\usepackage[colorlinks=true, allcolors=blue]{hyperref}
\usepackage{booktabs}
\usepackage{longtable}
\usepackage{array}
\usepackage{lineno}

\begin{document}

\begin{frontmatter}
\title{Estimating Climate Sensitivity Using Bayesian Model Averaging for CMIP Models}
\runtitle{Estimating Climate Sensitivity Using BMA}

\begin{aug}
\author[Jiang]{\fnms{Jitong}~\snm{Jiang}\textsuperscript{*}\ead[label=e1]{jitong.jiang@emory.edu}}
\author[Shi]{\fnms{Skylar}~\snm{Shi}\textsuperscript{*}\ead[label=e2]{shi.2154@buckeyemail.osu.edu}}
\author[Raftery]{\fnms{Adrian E.}~\snm{Raftery}\textsuperscript{\dag}\ead[label=e3]{raftery@uw.edu}}

\address[Jiang]{Department of Biostatistics and Bioinformatics, Emory University.\\ \printead{e1}}
\address[Shi]{Department of Statistics, The Ohio State University.\\ \printead{e2}}
\address[Raftery]{Department of Statistics, University of Washington.\\ \printead{e3}}

\footnotetext[1]{These authors contributed to this paper equally.}
\footnotetext[2]{Corresponding author.}

\end{aug}
\begin{abstract}
The Transient Climate Response to cumulative \ce{CO2} Emissions (TCRE) is a key metric for linking greenhouse gas emissions to global temperature change and informing climate policy. However, extant estimates of the TCRE often depend on subjective model selection or assumed sensitivity ranges, with limited validation against observed data. We develop a fully statistical data-driven approach using a Bayesian Model Averaging (BMA) approach to estimate the TCRE. This uses 37 climate models from the  Coupled Model Intercomparison Project phase 6 (CMIP6), weighted according to their consistency with observed temperature data. Compared to the Intergovernmental Panel on Climate Change (IPCC)’s Sixth Assessment Report (AR6) TCRE estimate, our BMA approach yields a very likely range (90\% interval) that overlaps substantially with that from the AR6, but with a higher mean and a lower standard deviation.  The resulting projections of global temperature change to 2100 show somewhat higher warming and less uncertainty than some current methods. The use of statistical modeling methods makes it easier to validate the approach and to partition the uncertainty according to its sources. Out-of-sample predictive validation shows the method to be well calibrated. Variance decomposition shows model parameter uncertainty to be a main source of projection variance.
\end{abstract}

\end{frontmatter}

\section{Introduction}
The Coupled Model Intercomparison Project (CMIP) is a collaborative global effort initiated by the World Climate Research Programme (WCRP) in 1995 to provide climate predictions that enhance our understanding of past and future climate change \citep{Meehl2000}. The CMIP integrates climate model experiments conducted by various modeling groups worldwide to enhance climate models and support national and international climate change assessments \citep{Meehl2000}. Every five to six years, the CMIP is updated to account for new data and methodological improvements.

Data and results from the various CMIP phases are widely used in the assessment reports of the Intergovernmental Panel on Climate Change (IPCC). These reports form the scientific foundation for global climate policy, offering scientific guidance and policy recommendations on climate change to governments and international organizations (\citealp{howarth2016exploring}, \citealp{pielke1998rethinking}). Among these, the CMIP Phase 5 (CMIP5) and Phase 6 (CMIP6) are the most recent and widely used generations of models. Based on the CMIP's simulation data, the IPCC provides predictions of future climate changes, aiding countries in setting emission reduction targets and adaptation strategies and supporting international climate negotiations (\citealp{rosenzweig2017assessing}, \citealp{chen2020quantifying}).

In the IPCC's Fifth Assessment Report (AR5), four Representative Concentration Pathways (RCPs) were used to explore future greenhouse gas emission scenarios, ranging from stringent mitigation (RCP2.6) to high emissions (RCP8.5) \citep{van2011representative}. For the AR6, these were updated to a new set of scenarios known as Shared Socioeconomic Pathways (SSPs), first released in 2016 \citep{bauer2017shared,kriegler2010socio,rao2019income}. These five scenarios were adopted in CMIP6, replacing the four RCPs from CMIP5 with updated pathways, namely SSP1-1.9, SSP1-2.6, SSP2-4.5, SSP3-7.0, and SSP5-8.5.   The outputs of CMIP models have also been used in regression-based analyses to estimate long-term climate outcomes. For example, \citet{liu2021country} used CMIP model results to regress global temperature on cumulative \ce{CO2}  emissions and averaged across models to generate probabilistic projections of future temperature.

One key metric derived from CMIP models is climate sensitivity, which quantifies the climate system's response to changes in greenhouse gas concentrations, typically measured as the change in global average temperature per unit of emitted \ce{CO2} (\citealp{hansen2013climate}, \citealp{stainforth2005uncertainty}). Climate sensitivity is traditionally measured in two ways: the Equilibrium Climate Sensitivity (ECS), which represents the long-term temperature response to a doubling of atmospheric \ce{CO2}, and the Transient Climate Response (TCR), which captures the temperature change at the time of \ce{CO2} doubling under a gradual increase scenario. While these metrics provide valuable insights into climate dynamics, they are limited in directly linking temperature change to cumulative \ce{CO2} emissions, which is a more relevant measure for global warming projection.


Thus a more relevant metric is the Transient Climate Response to cumulative \ce{CO2} Emissions (TCRE), defined as the globally averaged surface temperature change per unit of cumulative \ce{CO2} emissions \citep{macdougall2016transient}. Unlike the ECS and the TCR, which depend on emission pathways and climate feedbacks over different timescales, TCRE exhibits a near-linear relationship with cumulative emissions, making it a robust tool for climate projection and mitigation planning \citep{macdougall2015origin}. This near-linear relationship arises from the approximate balance between radiative forcing and ocean heat uptake, which leads to a proportional increase in global mean temperature with cumulative \ce{CO2} emissions. 

The IPCC’s Sixth Assessment Report \citep[AR6]{masson2021climate} adopted an approach to estimating TCRE that used both expert judgement and statistical analysis, based on multiple lines of evidence. It used the decomposition of the TCRE into two factors, the TCR and the airborne fraction (AF) (AR6, Section 5.5.1.3) \citep{Allen2009,Matthews2009}. The TCR was assessed by expert judgement using multiple lines of evidence (AR6, Section 7.5.5). The AF was assessed by expert judgement from CMIP6 models (AR6, Section 5.5.1.4) \citep{Arora2020,JonesFriedlingstein2020}. In this approach, some of the CMIP6 models were excluded because of being judged to be extreme, but were otherwise equally weighted. This led to an estimate of TCRE of 0.45$^{\circ}$C/1000 GtCO$_2$, with very likely range (90\% interval) 0.27–0.63. This approach has some limitations. Its reliance on subjective expert opinion makes it hard to assess in conventional statistical terms. This also makes it hard to validate quantitatively. It further makes it difficult to partition the uncertainty into its sources.

Here we take a fully statistical approach that weights the CMIP6 models based on their performance relative to observed data, using out-of-sample forecast validation, implemented through Bayesian model averaging (BMA). This reduces the dependence on subjective expert judgement, making the process more transparent and reproducible. It also  makes it easier to assess uncertainty in the forecasts using standard statistical tools. Finally, it allows one to partition the resulting uncertainty about TCRE into its components, using a standard statistical approach based on the analysis of variance.

To estimate TCRE, it is important to account for differences in model performance and their consistency with observed climate data. While poorly performing models may still exhibit a plausible relationship between warming and emissions, treating all models equally ignores heterogeneity in their ability to reproduce historical climate patterns. This equal-weighting assumption may reduce the reliability of the resulting estimates. Weighting provides a data-driven approach to address this issue by assigning greater influence to models that better align with observations while downweighting those with larger discrepancies.

Existing climate model weighting and selection methods vary in their assumptions and criteria.  \cite{sharmila2015future} and \cite{mahony2022global} selected models based on criteria such as maximum and minimum temperature and climatological seasonal mean. However, these selection approaches rely on subjective judgment.  Other approaches rely on the ECS and the TCR as key selection criteria. \cite{huard2022estimating} estimated the probability of RCP and SSP scenarios using Bayesian inference applied to probabilistic emissions from Integrated Assessment Models (IAMs). They first derived the compatible emissions required to produce the prescribed CO\textsubscript{2} concentrations in CMIP models. These compatible emissions were then compared with probabilistic IAM-generated emissions, and Bayes' theorem was applied to infer the likelihood of each scenario based on how well the IAM emissions align with the required emissions for each RCP/SSP.

To evaluate the efficiency of the selection methods, \cite{boyles2024approaches} explored different approaches for addressing the ``hot model'' problem in CMIP6, where some models exhibit excessively high ECS and TCR, potentially leading to overestimated warming projections. They evaluated four methods: retaining all models (\citealp{bureau2011west}, \citealp{lawrence2021divergent}), filtering models based on IPCC AR6-assessed ECS and TCR ranges (\citealp{tokarska2020past}, \citealp{ribes2021making}, \citealp{hausfather2022climate}), applying Bayesian Model Averaging (BMA) to probabilistically adjust model weights \citep{massoud2023bayesian}, and using Global Warming Levels (GWLs) to analyze impacts at specific warming thresholds \citep{ipcc2021warming}. Their findings suggest that BMA offers a probabilistic framework that preserves ensemble diversity, but is computationally intensive. ECS and TCR screening provides a straightforward filtering mechanism but may discard useful models. 

 \cite{wootten2023assessing} explored the sensitivity of climate model weighting methods by implementing and comparing five different approaches. The \textit{unweighted model mean} serves as a baseline, treating all models equally without assigning specific weights. \textit{Historical skill weighting (Skill)} weights models based on their historical performance, using root mean squared error (RMSE) to assess their ability to simulate past climate conditions (\citealp{wootten2020effect}, \citealp{massoud2020bayesian}). \textit{Historical skill and historical independence weighting (SI-h)} refines Skill weighting by incorporating model dependence using an inter-model distance matrix, reducing the influence of highly correlated models \citep{sanderson2017model}. \textit{Historical skill and future independence weighting (SI-c)}, extends SI-h by incorporating future climate signal-based independence, ensuring that models with similar future projections do not dominate the ensemble \citep{wootten2020effect}.

 BMA averages the predictive distributions from different models according to weights that are proportional to their Bayesian posterior model probabilities \citep{MadiganRaftery1994,Hoeting1998,Hoeting1999}. These in turn reflect how well the models match observed data. The BMA framework was first applied to atmospheric models in the weather forecasting context \citep{Raftery2005}.
 The approach was later applied to CMIP climate models using a Markov Chain Monte Carlo (MCMC) sampling approach to optimize model weights while considering uncertainty and observational consistency  (\citealp{massoud2019global}, \citealp{massoud2020bayesian}). By comparing results using both CMIP5 GCMs and their downscaled projections created with the Localized Canonical Analogs counterparts,  they found that BMA was the most effective at reducing uncertainty, while its results were also the most sensitive to input assumptions.

\cite{massoud2023bayesian} applied BMA to adjust model weights, ensuring alignment with the IPCC AR6-assessed likely range of ECS (2.5–4.0 \degree C) and TCR (1.4–2.2 \degree C). They evaluated thousands of weight combinations and optimized them to best match these target ECS and TCR distributions, downweighting models that deviate significantly while preserving ensemble diversity. However, their weighting criteria are still based on the IPCC’s predefined climate sensitivity estimates rather than direct model performance. In contrast, here we implement BMA by assigning weights based on each model’s performance against actual observational data, yielding a data-driven approach to model weighting. 

\cite{li2024constrain} introduced a composite likelihood approach to estimate ECS by integrating Energy Balance Models with CMIP5 and CMIP6 simulations. Instead of directly using BMA or emergent constraints, they applied Bayesian composite likelihood estimation to fit Energy Balance Models to surface temperature and top-of-atmosphere heat flux responses from a quadrupled CO\textsubscript{2} forcing experiment across 31 climate models. The method assigns data-driven weights to models based on their consistency with a common ECS estimate, downweighting models with extreme or unrealistic responses. Their approach produced a constrained ECS estimate of 3.3\degree C (95\% CI: 3.18\degree C–3.44\degree C), much narrower than the IPCC AR6 likely range (2.5\degree C–4\degree C). 


While existing model selection and weighting methods provide valuable insights, they suffer from limitations, including reliance on subjective filtering criteria, computational inefficiency, and a lack of direct observational validation. Many approaches prioritize consistency with predefined ECS and TCR distributions rather than optimizing model selection based on empirical performance. Our study seeks to address these issues by implementing a data-driven BMA approach that directly evaluates models against observational data,  with the aim of improving transparency, robustness, and reliability in climate projections.

The structure of the paper is as follows. In Section 2, we describe the data sources, variable construction, and CMIP6 model grouping, along with the two model selection strategies used in the analysis. In Section 3, we present the Bayesian Model Averaging framework, including the autoregressive formulation and estimation of posterior model probabilities. Section 4 reports the results of model averaging, projections under various emission scenarios, variance decomposition, and validation. Section 5 discusses the main conclusions, implications of our results, and possible directions for future work. Additional derivations and technical details are provided in the Supplementary Material.

\section{Data}
\subsection*{CO$_2$ and temperature data}

Our approach follows a “train on simulations, weight by observations” strategy. Specifically, model training is based on CMIP6 simulation data, denoted by \( \tilde{D} \). In our analysis, CMIP6 provides trajectories of cumulative CO$_2$ emissions as the explanatory variable, along with simulated temperature anomalies as the response, forming the basis for model estimation.

CMIP6 temperature data were obtained from the KNMI Climate Explorer platform \citep{knmi_climate_explorer}, while cumulative CO$_2$ emissions were derived from SSPs, which provide distinct CO$_2$ emissions trajectories beginning in 2015 (\citealp{Riahi2017, Gidden2018}). Since our analysis is based on cumulative CO$_2$ emissions, we start from 2016 to ensure that cumulative quantities are well-defined and consistently aligned with the corresponding temperature responses. Therefore, the analysis focuses on the period 2016--2024. We consider four commonly used scenarios (SSP1-2.6, SSP2-4.5, SSP3-7.0, SSP5-8.5), corresponding to the standard range of future emissions pathways. The first SSP, SSP1-1.9, is omitted since it corresponds to unrealistically low future emissions. The simulated data, \( \tilde{D} \), are used to estimate model-specific parameters governing the temperature response.

To evaluate model performance, we use observational temperature data, denoted by \( D\) . In this study, ``temperature" refers to temperature anomaly, defined as the deviation of observed global temperatures from a baseline period, specifically calculated relative to the 1961-1990 average in the HadCRUT5 dataset \citep{Morice2021}. These data serve as a reality-based benchmark for assessing how well each model reproduces observed temperature patterns. By comparing model-implied temperature responses with observed temperature anomalies, we assess model performance, which is later used to construct model weights in the Bayesian Model Averaging framework.

\subsection*{CMIP model selection}

The CMIP6 temperature data contain 51 realizations from multiple climate models. Since several models provide multiple realizations (e.g., ensemble members), treating each realization equally would lead to over-representation of models with larger ensemble sizes. To address this, we grouped realizations into 37 model families based on their base model names by removing realization-specific suffixes (See Table~\ref{tab:model_map} for details).

We then considered two model selection strategies: 

(1) Random realization sampling. In this strategy, for each model family, we randomly selected one realization using uniform sampling, ensuring that each model family contributes exactly one trajectory to the analysis. This yields a dataset of 37 realizations, each one corresponding to a distinct model family. All subsequent model estimation and inference are performed using this sampled set. 

(2) All-realization aggregation. In this alternative strategy, we retained all 51 realizations and performed model estimation separately for each realization, resulting in a set of realization-level weights that sum to one. To obtain model-family-level weights, we grouped realizations according to their model family and computed the average weight within each family. Since the resulting family-level weights no longer sum to one, we applied a normalization step to rescale them so that the total weight across the 37 model families equals one. This procedure ensures that model families with multiple realizations do not receive disproportionately large total weight, while still incorporating within-family variability through the initial realization-level weighting.

Results based on the random sampling approach are presented in the main text. The aggregation-based results are reported in the Supplementary Material and are qualitatively consistent with the main findings.

{\small
\setlength{\tabcolsep}{3pt}
\renewcommand{\arraystretch}{1.1}

\begin{longtable}{lll}
\caption{CMIP model selection: 51 ensemble members grouped into 37 climate models.}
\label{tab:model_map} \\
\toprule
Model ID & Climate Model & Ensemble Members \\
\midrule
\endhead
\bottomrule
\endfoot
1  & ACCESS-CM2      & ACCESS-CM2 \\
2  & ACCESS-ESM1-5   & ACCESS-ESM1-5 \\
3  & AWI-CM-1-1-MR   & AWI-CM-1-1-MR \\
4  & BCC-CSM2-MR     & BCC-CSM2-MR \\
5  & CAMS-CSM1-0     & CAMS-CSM1-0 \\
6  & CanESM5         & CanESM5, CanESM5-p1, CanESM5-p2 \\
7  & CanESM5-CanOE   & CanESM5-CanOE, CanESM5-CanOE-p2 \\
8  & CESM2           & CESM2 \\
9  & CESM2-WACCM     & CESM2-WACCM \\
10 & CIESM           & CIESM \\
11 & CMCC-CM2-SR5    & CMCC-CM2-SR5 \\
12 & CNRM-CM6-1      & CNRM-CM6-1, CNRM-CM6-1-f2 \\
13 & CNRM-CM6-1-HR   & CNRM-CM6-1-HR, CNRM-CM6-1-HR-f2 \\
14 & CNRM-ESM2-1     & CNRM-ESM2-1, CNRM-ESM2-1-f2 \\
15 & EC-Earth3       & EC-Earth3 \\
16 & EC-Earth3-Veg   & EC-Earth3-Veg \\
17 & FGOALS-f3-L     & FGOALS-f3-L \\
18 & FGOALS-g3       & FGOALS-g3 \\
19 & FIO-ESM-2-0     & FIO-ESM-2-0 \\
20 & GFDL-ESM4       & GFDL-ESM4 \\
21 & GISS-E2-1-G     & GISS-E2-1-G, GISS-E2-1-G-f2, GISS-E2-1-G-p1, GISS-E2-1-G-p3 \\
22 & HadGEM3-GC31-LL & HadGEM3-GC31-LL, HadGEM3-GC31-LL-f3 \\
23 & HadGEM3-GC31-MM & HadGEM3-GC31-MM, HadGEM3-GC31-MM-f3 \\
24 & INM-CM4-8       & INM-CM4-8 \\
25 & INM-CM5-0       & INM-CM5-0 \\
26 & IPSL-CM6A-LR    & IPSL-CM6A-LR \\
27 & KACE-1-0-G      & KACE-1-0-G \\
28 & MCM-UA-1-0      & MCM-UA-1-0, MCM-UA-1-0-f2 \\
29 & MIROC-ES2L      & MIROC-ES2L, MIROC-ES2L-f2 \\
30 & MIROC6          & MIROC6 \\
31 & MPI-ESM1-2-HR   & MPI-ESM1-2-HR \\
32 & MPI-ESM1-2-LR   & MPI-ESM1-2-LR \\
33 & MRI-ESM2-0      & MRI-ESM2-0 \\
34 & NESM3           & NESM3 \\
35 & NorESM2-LM      & NorESM2-LM \\
36 & NorESM2-MM      & NorESM2-MM \\
37 & UKESM1-0-LL     & UKESM1-0-LL, UKESM1-0-LL-f2 \\

\end{longtable}
}

\section{Methodology}

We use Bayesian Model Averaging to estimate climate sensitivity using simulations from the CMIP. The primary objective is to infer the posterior distribution of the global temperature response to cumulative \(\mathrm{CO_2}\) emissions, integrating information across multiple climate models while weighting them according to their alignment with historical observations.

We model the global mean surface temperature, \( y_t \), in year \( t \) as a function of cumulative \(\mathrm{CO_2}\) emissions, \( x_t \), incorporating an intercept term, \( \alpha \), and a slope, \( \beta \). Here, $\beta$ corresponds to the TCRE, defined as the change in global mean temperature per unit increase in cumulative CO$_2$ emissions. Due to the temporal dependence in the data, we specify the model as a first-order autoregressive (AR(1)) process, such that the temperature follows
\begin{equation}
    y_t = \alpha + \beta x_t + u_t.
    \label{eq:basic_model}
\end{equation}
Here \( u_t \) is an autocorrelated error term modeled as $u_t = \phi u_{t-1} + \varepsilon_t, \varepsilon_t \sim \mathcal{N}(0, \sigma^2),$ where \( \phi \) represents the autocorrelation parameter of the error term, and \( \varepsilon_t \) is an independent 
and identically distributed (iid) normal error with variance \( \sigma^2 \). The linear specification is motivated by the well-established TCRE relationship, which implies an approximately proportional relationship between temperature and cumulative \ce{CO2} emissions. The residual term is modeled as an AR(1) process to account for temporal dependence in temperature anomalies. Such persistence is expected due to the thermal inertia of the climate system, particularly ocean heat storage, and has been widely documented in climate time series.

Assume that CMIP consists of \( K \) distinct climate models, each providing different estimates of the parameters \(\alpha\) and \(\beta\) when fitting the AR(1) model. Let \(\alpha_k\) and \(\beta_k\) denote the corresponding parameters for the \( k \)-th climate model, \( M_k \). Our objective is to determine the posterior distribution of \(\alpha\) and \(\beta\) within the BMA framework, which is expressed as  

\begin{equation}
    p(\alpha, \beta  \mid D, \tilde{D}) = \sum_{k=1}^{K} p(\alpha, \beta \mid \tilde{D}, M_k) p(M_k \mid  D),
    \label{eq:post_prob}
\end{equation}
 where \( D \) represents the observational data from HadCrut5, \( \tilde{D} \) denotes the model-simulated data from the climate models, and \( M_k \) corresponds to the \( k \)-th climate model. The term \( p(M_k | D) \) represents the posterior probability of model \( M_k \), which quantifies the model’s consistency with observed data and serves as a weight in the BMA framework.  

Thus, the procedure can be summarized in the following steps:

(1) Estimate the parameters $(\alpha_k, \beta_k, \phi_k)$ for each climate model using the model-simulated data $\tilde{D}$;

(2) Compute the posterior model probabilities $p(M_k \mid D)$ based on observational data;

(3) Obtain the posterior distribution of $(\alpha, \beta)$ via Bayesian model averaging;

(4) Construct the predictive distribution of future temperature change.

\subsection*{Estimating $\alpha$, $\beta$, $\phi$  from model $k$}

While $\alpha$, $\beta$ and $\phi$ have the same interpretation for each climate model, here for clarity we add a subscript $k$ to each to indicate posterior distributions based on model $k$.  For each climate model \( M_k \), we denote the simulated global mean temperature in the CMIP dataset by \( \tilde{y}_{t,k,s} \), representing the temperature in year \( t \) under scenario \( s \). We assume that the parameters $\alpha_k$ and $\beta_k$ are shared across scenarios for a given climate model $M_k$. This is motivated by the widely observed near-linear relationship between global temperature and cumulative CO$_2$ emissions, which is approximately invariant across emission scenarios. Under this assumption, different scenarios contribute additional information on the same underlying temperature–emissions relationship, improving estimation efficiency. Then the relationship between temperature and cumulative \(\mathrm{CO_2}\) emissions follows the regression model: 

\begin{equation}
\tilde{y}_{t,k,s} = \alpha_k + \beta_k \tilde{x}_{t,s} + u_{t,k,s},
\end{equation}
where \( \tilde{x}_{t,s} \) represents the cumulative \(\mathrm{CO_2}\) emissions at time \( t \) for scenario \( s \), as provided by the CMIP. 
The error term \( u_{t,k,s} \) accounts for residual variability and exhibits autocorrelation, which we model using an AR(1) process: $u_{t,k,s} = \phi_k u_{t-1,k,s} + \varepsilon_{t,k,s}, \ \varepsilon_{t,k,s} \overset{\text{iid}}{\sim} \mathcal{N}(0, \sigma_k^2).$ The innovation term \( \varepsilon_{t,k,s} \) is iid across \( t \) for each model \( k \), but the variance \( \sigma_k^2 \) varies between models, reflecting model-specific uncertainty.  

To account for autocorrelation in the residuals and obtain unbiased estimates of \( \alpha_k \) and \( \beta_k \), we use a Cochrane-Orcutt estimation method \citep{CochraneOrcutt1949}. To do this, we apply a transformation to remove serial dependence, namely: 
\begin{equation}
    \tilde{y}'_{t,k,s} = \tilde{y}_{t,k,s} - \phi_k \tilde{y}_{t-1,k,s}, \quad
    \tilde{x}'_{t,s} = \tilde{x}_{t,s} - \phi_k \tilde{x}_{t-1,s}, \quad
    \alpha'_k = \alpha_k (1 - \phi_k), \quad
    \beta'_k = \beta_k.
\end{equation}  
This transformation effectively removes the temporal correlation in the residuals. The resulting transformed regression model is then given by:  
\begin{equation}
    \tilde{y}'_{t,k,s} = \alpha'_k + \beta'_k \tilde{x}'_{t,s} + \varepsilon_{t,k,s}.
    \label{eq:trans_model}
\end{equation}  
This formulation ensures that the ordinary least squares (OLS) estimates of \( \alpha_k \) and \( \beta_k \) remain unbiased, while accounting for the autoregressive structure of the residuals.

For each climate model \( M_k \) ($k =1,\ldots,37$), the time index extends from \( t = 1 \) to \( T \), and the model incorporates \( S_k \) different emissions scenarios. Given a specified value of \(\phi_k\), the transformed parameters \((\alpha'_k, \beta'_k)\) can be estimated using OLS based on the \( T \times S_k \) data points. Once \((\alpha'_k, \beta'_k)\) are obtained, the residuals \( u_{t,k,s} \) can be computed, and \(\phi_k\) is then estimated as the correlation between \( u_{t,k,s} \) and \( u_{t-1,k,s} \) for \( t = 2, \dots, T \) and \( s = 1, \dots, S_k \). Since all climate models share the same temporal structure, \( T \) is constant across all \( M_k \). However, the number of emissions scenarios, \( S_k \), varies across models, taking values from the set \( \{1,2,3,4\} \).

Using this framework, we estimate the parameters \((\alpha_k, \beta_k, \phi_k)\) for each model \( M_k \) through the following iterative procedure:
\begin{enumerate}
    \item Initialize $\phi_k = 0.5$.
    \item Estimate $(\alpha'_k, \beta'_k)$ by OLS given $\phi_k$ by OLS from \eqref{eq:trans_model}. This gives $\alpha_k = \alpha'_k / (1 - \phi_k)$, $\beta_k = \beta'_k$.
    \item Form $u_{t,k,s} = \tilde{y}_{t,k,s} - (\alpha_k + \beta_k \tilde{x}_{t,s})$.
    \item Estimate $\phi_k$ from $u_t = \phi u_{t-1} + \varepsilon_t$ as the correlation between $u_{t,k,s}$ and $u_{t-1,k,s}$ for $t = 2, \ldots, T$ and $s = 1, \ldots, S_k$. 
    \item If $\beta_k$ barely changes from the last iteration (change is less than 1\%), estimate $\text{Var}(\beta_k)$ as the OLS variance from the last iteration. Stop. We monitor convergence using $\beta$, as it is the primary parameter of interest (climate sensitivity). The intercept $\alpha$ is re-estimated jointly with $\beta$ at each iteration via OLS and stabilizes once $\beta$ converges. 
    \item If $\beta_k$ did change from the last iteration, return to step 2.
\end{enumerate}
We denote by $\hat{\alpha}_k$, $\hat{\beta}_k$ and $\hat{\phi}_k$ the estimates of $\alpha_k$, $\beta_k$ and $\phi_k$ from this process, with covariance matrix $\Sigma_k$ of $\hat{\alpha}_k$ and $\hat{\beta}_k$. A simpler two-step approach based on ordinary least squares followed by estimation of $\phi$ is also possible; however, it would not fully account for serial correlation in the estimation of regression coefficients. The iterative procedure used here corresponds approximately to a generalized least squares approach.

\subsection*{Finding posterior model probabilities}

To compute the posterior model probabilities \( p(M_k | D) \), we use the observed data \( D \). 
The observational data follow a regression model analogous to the one used for the CMIP simulations. In the case of CMIP6, the time range is from 2016 to 2024, consisting of \( n = 9 \) years. The regression model is specified as $y_t = \hat{\alpha}_k + \hat{\beta}_k x_t + u_{t,k}$ where the residual term follows $u_{t,k} = \hat{\phi}_k u_{t-1,k} + \varepsilon_{t,k}$. Here, the parameters \( \hat{\alpha}_k \), \( \hat{\beta}_k \), and \( \hat{\phi}_k \) are fixed at their previously estimated values obtained from the CMIP model-simulated data. The residual variance for each model \( M_k \) is then estimated as $\hat{\sigma}_k^2 = \frac{1}{n} \sum_{t} \varepsilon_{t,k}^2.$ For CMIP6 this reduces to $\hat{\sigma}^2_k = \frac{1}{9} \sum_{t=2016}^{2024} \varepsilon_{t,k}^2$ Following the Bayesian Information Criterion (BIC)-based  approach outlined by \cite{raftery1995bayesian}, the posterior probability of model \( M_k \) is given by $p(M_k | D) \propto (\hat{\sigma}_k^2)^{-n/2}.$ The normalized model weight for model $M_k$ is then $w_k = \frac{(\hat{\sigma}_k^2)^{-n/2}}{\sum_{k=1}^{K} (\hat{\sigma}_k^2)^{-n/2}}.$

\subsection*{Finding the posterior joint distribution of $\alpha$, $\beta$}
We approximate the posterior distribution \( p \left( \alpha, \beta \mid \tilde{\mathcal{D}}, M_k \right) \) for each climate model \( M_k \) by a normal distribution with mean $\begin{bmatrix} \hat{\alpha}_k \\ \hat{\beta}_k \end{bmatrix}$ and covariance matrix \( \Sigma_k \). Thus, based on Equation~\eqref{eq:post_prob}, the posterior distribution of \( (\alpha, \beta) \) under the BMA framework is given by:  
\begin{equation}
p \left( \begin{bmatrix} \alpha \\ \beta \end{bmatrix} \mid \mathcal{D, \tilde D} \right) = \sum_{k=1}^{K} w_k \mathcal{N} \left( \begin{bmatrix} \hat{\alpha}_k \\ \hat{\beta}_k \end{bmatrix}, \Sigma_k\right).
\end{equation}
 We approximate this mixture distribution with a single normal distribution whose mean is given by the weighted average of the individual model estimates, $\begin{bmatrix} \hat{\alpha} \\ \hat{\beta} \end{bmatrix} = \begin{bmatrix} \sum_{k=1}^{K} w_k \hat{\alpha}_k \\ \sum_{k=1}^{K} w_k \hat{\beta}_k \end{bmatrix}.$ This approximation matches the first two moments of the mixture distribution and is used to facilitate analytical tractability in subsequent derivations. The corresponding posterior covariance matrix \( \Sigma \) is derived using the Law of Total Variance and the Law of Total Covariance, yielding: $\Sigma_{1,1} = \sum_{k=1}^{K}w_k\Sigma_{k,1,1} + \text{Var}(\hat{\alpha}_{k}), $ $\Sigma_{2,2} = \sum_{k=1}^{K}w_k\Sigma_{k,2,2} + \text{Var}(\hat{\beta}_{k}),$ and $\Sigma_{1,2} = \Sigma_{2,1} = \sum_{k=1}^{K}w_k\Sigma_{k,1,2} + \text{Cov}(\hat{\alpha}_{k},\hat{\beta}_{k}).$ Alternative approaches, such as direct sampling from the mixture distribution, are possible but would substantially increase computational complexity without materially affecting the results.



\begin{figure}
    \centering
    \includegraphics[width=0.8\linewidth]{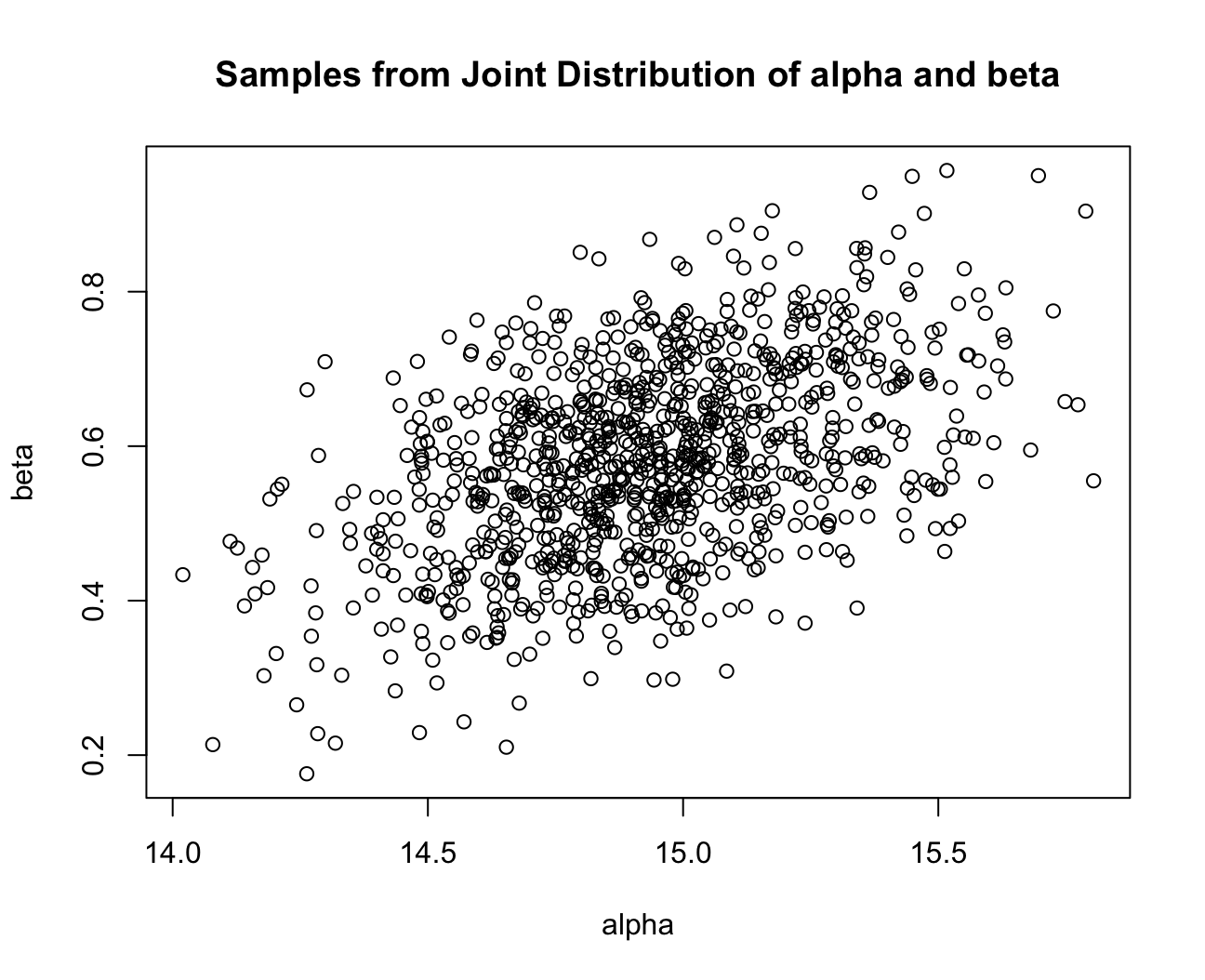}
    \caption{Samples from the joint posterior distribution of $\alpha$ and $\beta$ obtained from the BMA model. The scatterplot reveals positive dependence between the two parameters, consistent with the estimated posterior correlation of 0.44.}
    \label{fig:alpha_beta}
\end{figure}

\subsection*{Predictive distribution of future temperature change}
We define the global mean surface temperature and cumulative \(\mathrm{CO_2}\) emissions at year \( t \) as \( y_{t} \) and \( x_{t} \), respectively. The baseline year is set to 1850, with \( u_{1850} = 0 \), consistent with the IPCC definition of the pre-industrial period as preceding significant human influence on atmospheric composition.

Given the posterior joint distribution $\begin{bmatrix} \alpha \\ \beta \end{bmatrix} \mid \mathcal{D}, \tilde{\mathcal{D}} \sim \mathcal{N} \left( \begin{bmatrix} \hat{\alpha} \\ \hat{\beta} \end{bmatrix}, \Sigma \right)$,
and the AR(1) temperature model in Equation~\eqref{eq:basic_model}, the temperature change $\Delta T$ over a prediction horizon of $m$ years, from year $t_1$ to year $t_2 = t_1 + m$, is given by:

\begin{equation}
    \Delta T = y_{t_2} - y_{t_1} = \beta \Delta x + \Delta u , 
\end{equation}
where \( \Delta x = x_{t_2} - x_{t_1} \) represents the change in cumulative \(\mathrm{CO_2}\) emissions from \( t_1 \) to \( t_2 \), and \( \Delta u = u_{t_2} - u_{t_1} \) corresponds to the change in the autocorrelated error term.

We approximate the distribution of \( \Delta T \) by a Normal distribution with mean and variance:

\begin{align}
\left\{
\begin{array}{l}
\mathbb{E}[\Delta T] = \Delta x \hat{\beta}, \\
\text{Var} (\Delta T) = \mathbb{E}_{\phi,\sigma^2} \left[\sigma^2 (\phi^m - 1)^2 \times \frac{1 - \phi^{2t_1}}{1 - \phi^2} + \sigma^2 \times \frac{1 - \phi^{2m}}{1 - \phi^2} \right] 
+ \Delta x^2 \Sigma_{2,2}.
\end{array}
\right.
\label{eq:temp_dis}
\end{align}
These values are derived in Supplementary Material, Section~S1.1.

Since the parameters \( \phi \) and \( \sigma^2 \) are not directly observable, we estimate \( \text{Var}(\Delta T) \) using a Monte Carlo sampling approach. We draw 1,000 samples of \( (\alpha_j, \beta_j) \) from  $\mathcal{N} \left( \begin{bmatrix} \hat{\alpha} \\ \hat{\beta} \end{bmatrix}, \Sigma \right),$ and for each sample, we fit the model to estimate the corresponding autoregressive parameter \( \phi_j \) and residual variance \( \sigma_j^2 \) (Additional implementation details are provided in Supplementary Material, Section~S1.3). The predictive variance, $\text{Var}(\Delta T)$, is then approximated by $\frac{1}{1000} \sum_{j=1}^{1000} \left[ \sigma_j^2 (\phi_j^m - 1)^2 \times \frac{1 - \phi_j^{2t_1}}{1 - \phi_j^2} + \sigma_j^2 \times \frac{1 - \phi_j^{2m}}{1 - \phi_j^2} \right] + \Delta x^2 \Sigma_{2,2}.$

\section{Results}

\subsection{Model averaging}
We began with an ensemble of 37 CMIP6 models. Applying BMA to these 37 models produced posterior mean estimates $\alpha = 14.93 \quad \text{and} \quad \beta = 0.58$ \degree C per 1000 Gt \(\mathrm{CO_2}\). The posterior standard deviations are 0.31 for $\alpha$ and $0.13$ for $\beta$. The posterior correlation between $\alpha$ and $\beta$ is 0.44 (see Figure \ref{fig:alpha_beta}). The posterior distribution of \(\beta\) can be compared with the IPCC AR6's Gaussian distribution of TCRE, which has mean 
\(0.45\) 
and standard deviation 
\(0.18\) per 1000 Gt \(\mathrm{CO_2}\). The very likely interval (90\% interval) from AR6 is thus 0.27--0.63, while from our approach it is 0.37--0.79. Thus our interval overlaps substantially with that from AR6, but the mean is somewhat higher and the uncertainty is somewhat smaller. Thus the BMA approach projects somewhat higher global mean temperature increases with lower uncertainty.

\begin{figure}[t]
    \centering
    \includegraphics[width=0.8\linewidth]{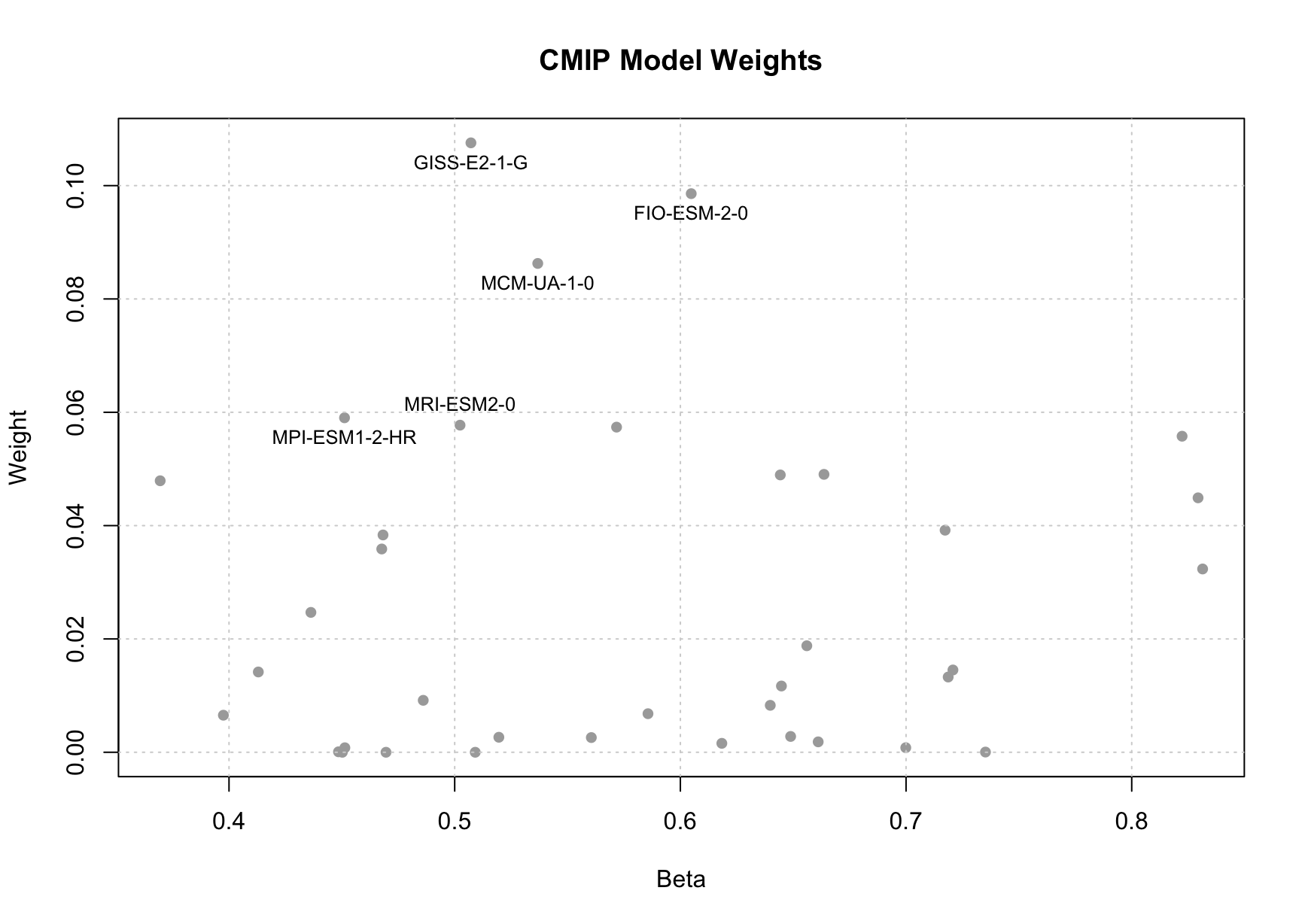}
    \caption{Posterior model weights from the BMA procedure applied to 37 CMIP6 climate models, where each weight reflects the model’s relative fit to observed historical temperatures. The regression coefficients $\beta_k$ (i.e., estimates of climate sensitivity) are estimated via OLS for each model.} 
    \label{weights}
\end{figure}

Figure~\ref{weights} displays the BMA weights for each of the 37 temperature models, which vary considerably. The model ``GISS-E2-1-G'' attains the highest weight (0.11), while other top contributors include ``FIO-ESM-2-0'' (0.10), ``MCM-UA-1-0'' (0.09), ``MPI-ESM1-1-HR'' (0.06), and ``MRI-ESM2-0''(0.06). A common characteristic of these high-weight models is that their \(\beta\) estimates lie near the median among all candidates, aligning with the notion that ``median-sensitivity'' models often carry significant influence under BMA. Some high-sensitivity models also receive moderate weights, indicating that their historical climate performance remains good enough to be somewhat favored in the posterior distribution. In contrast, seven models have weights close to zero, reflecting limited alignment with observed historical data. 

\subsection{Validation}

To assess the robustness and predictive accuracy of our BMA framework, we conducted two complementary validation experiments. The first evaluated the model's ability to recover historical temperature dynamics when trained on recent data, while the second examined its out-of-sample predictive performance.

Our BMA approach models temperature evolution as
\begin{equation}
y_t = y_{\text{start.year}} + \Delta T,
\end{equation}
where $\Delta T$ represents the predicted temperature change. The expected value of $\Delta T$ is given by
\begin{equation}
\mathbb{E}[\Delta T] = \Delta x \, \hat{\beta}, \quad \Delta x = x_t - x_{\text{start.year}},
\end{equation}
where $\Delta x$ denotes the cumulative increase in carbon emissions relative to the starting year.

The variance of $\Delta T$ incorporates both the posterior uncertainty in $\beta$ and additional sources of variability:
\begin{equation}
\mathrm{Var}(\Delta T) = \mathbb{E}_{\phi,\sigma^2} \left[
\sigma^2(\phi^m - 1)^2 \frac{1 - \phi^{2(\text{start.year}-1850)}}{1 - \phi^2}
+ \sigma^2 \frac{1 - \phi^{2m}}{1 - \phi^2}
\right]
+ \Delta x^2 \Sigma_{2,2},
\end{equation}
where $\Sigma_{2,2}$ is the posterior variance of $\beta$, and 1850 is taken as the pre-industrial baseline following IPCC conventions. The quantity  \( m = t - \text{start.year} \) represents the number of years since the model's starting point (i.e., the prediction horizon), and determines how uncertainty accumulates over time through the autoregressive structure.

\begin{figure}
    \centering
    \includegraphics[width=0.8\linewidth]{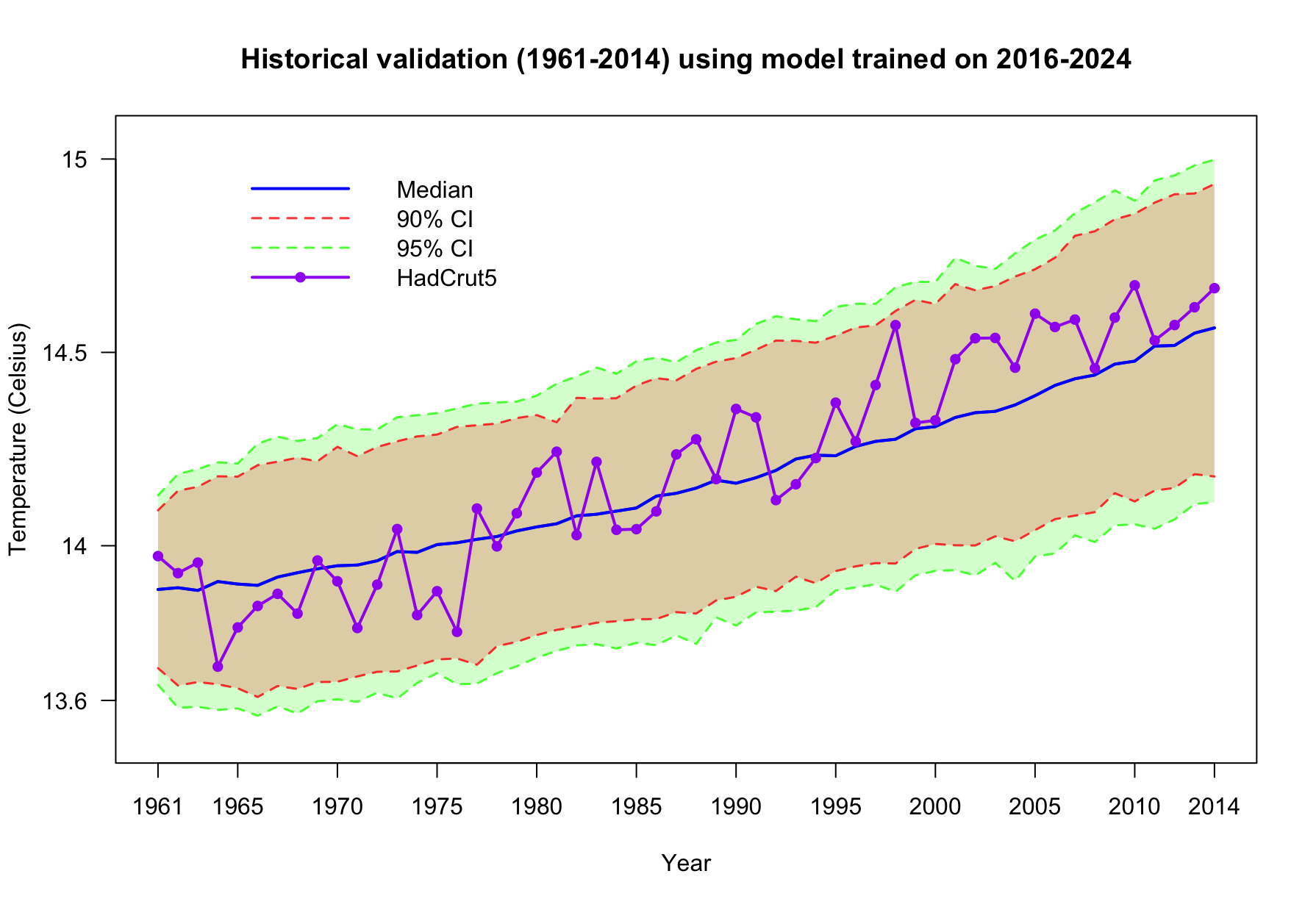}
    \caption{Historical validation (1961--2014) using the model trained on 2016--2024 data: posterior median (blue), 90\% and 95\% credible intervals (red and green dashed lines), and observed temperatures from HadCRUT5 (purple). The model captures the long-term trend and provides well-calibrated uncertainty.}
    \label{fig:his_validation}
\end{figure}

\begin{figure}[t!]
    \centering
\includegraphics[width=\linewidth]{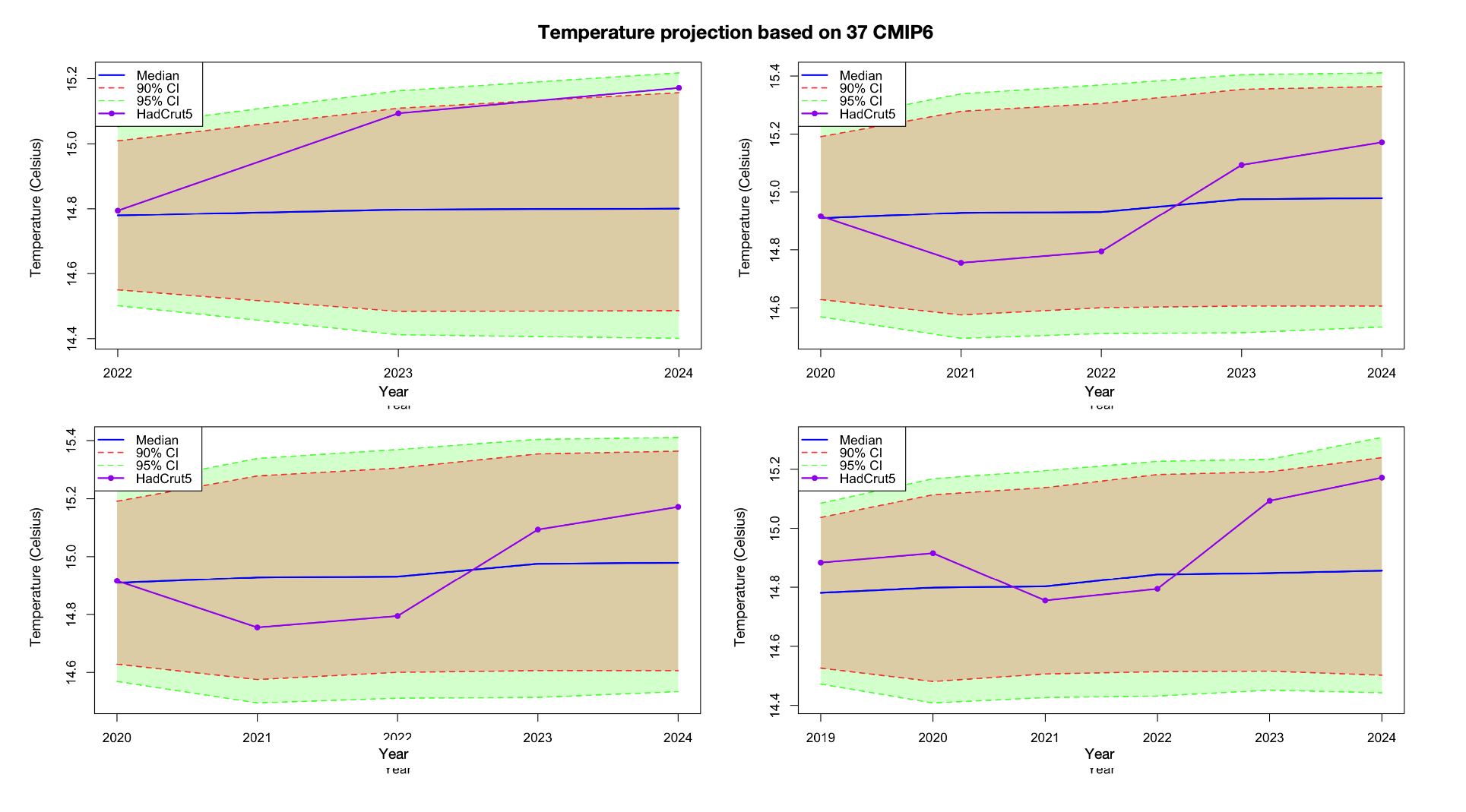}
    \caption{Out-of-sample validation of the BMA-based temperature projections using holdout periods ending in 2018, 2019, 2020, and 2021. The posterior median and 90\% and 95\% credible intervals are compared to observed global temperature anomalies from HadCRUT5.
}
    \label{fig:os_validation}
\end{figure}

\paragraph*{Historical validation}
In the first validation experiment, we assessed whether the model can recover historical temperature trajectories when trained on recent observations. Specifically, the BMA model was trained using data from 2016 to 2024, and then applied retrospectively to reconstruct global temperature evolution over the period 1960--2014. Temperature anomalies change were defined relative to the 1960, such that $\Delta T = T_t - T_{1960}$, and thus the effective validation period spanned 1961--2014. For each year in the validation period, we computed the posterior predictive median and credible intervals of temperature, and compared them with observed temperature anomalies from HadCRUT5. 

Figure \ref{fig:his_validation} presents the results. The model captured the long-term warming trend in historical data, and the observed temperatures were largely contained within the predictive intervals. Thus the BMA predictions of out of sample observations were well calibrated.

\paragraph*{Out-of-sample validation}
We further evaluated predictive performance using an out-of-sample validation experiment. Specifically, we conducted four out-of-sample validation tests using expanding training windows. Temperature anomalies change were relative to 2015. So in each test, the model was trained on temperature data from 2016 up to a cutoff year $\text{start.year} \in \{2018, 2019, 2020, 2021\}$, and then used to generate projections for the subsequent period through 2024. These projections were compared with observed temperature anomalies from HadCRUT5. Figure \ref{fig:os_validation} shows the posterior median and 90\% and 95\% credible intervals alongside observed temperatures. The results indicate that the BMA framework achieved good predictive accuracy, with the posterior median closely tracking observed trends. Moreover, the predictive intervals provided appropriate coverage, indicating well-calibrated uncertainty quantification Overall, the two validation experiments found the proposed BMA approach to provide accurate forecasts and well-calibrated uncertainty statements.


\subsection{Projection}

We generated 1{,}000 trajectories of \(\mathrm{CO_2}\) emissions spanning the period from 2025 to 2100 using a fully Bayesian hierarchical modeling framework, which integrates population growth, per-capita GDP, and carbon intensity \citep{raftery2017,liu2021country}. These trajectories correspond to projected CO$_2$ emissions and are used to construct cumulative emissions $x_t$, which serves as the explanatory variable in the temperature model. For each emissions trajectory, we drew samples from the temperature change distribution specified in Equation \eqref{eq:temp_dis} for each year, thereby constructing an ensemble of BMA-based temperature trajectories.  This allows us to combine information across models in a consistent way, accounting for differences in climate model behavior while producing an overall prediction with uncertainty. For comparison, we also created an ensemble of temperature trajectories using a benchmark TCRE-based implementation, in which the IPCC AR6 TCRE distribution is combined with the same linear temperature model.

\begin{figure}[t]
    \centering
\includegraphics[width=0.8\linewidth]{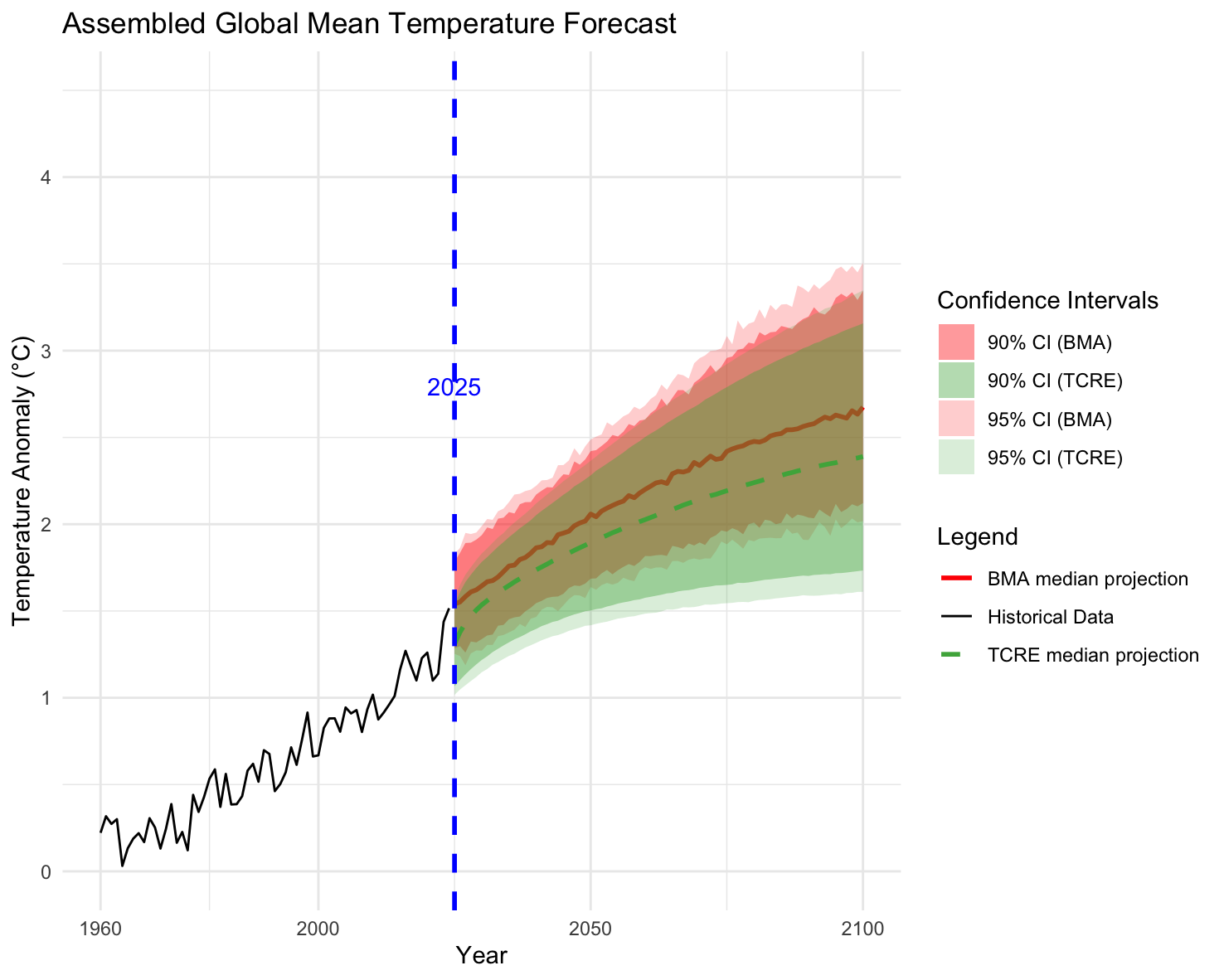}
    \caption{Projected global mean temperature anomalies from 2025 to 2100 using the BMA model (red) and the TCRE model (green). Shaded regions represent credible intervals: dark shading indicates the 90\% interval, and light shading indicates the 95\% interval. 
    }
    \label{projection_none}
\end{figure}

The BMA- and TCRE-based projections follow a generally similar trend over time (See Figure \ref{projection_none}). However, the BMA framework  yielded higher temperature projections and narrower predictive intervals. By 2100, the median projected temperature under the BMA approach reached \(2.67^\circ\mathrm{C}\) with a 90\% prediction interval of \((2.12, 3.34)\). For comparison, the TCRE method produced a median projection of \(2.39^\circ\mathrm{C}\) (90\% interval: \((1.73, 3.16)\)). Relative to the TCRE benchmark, the BMA projection reflected  higher expected warming and a narrower range of uncertainty. However, the two predictive intervals largely overlap, so the results from the two methods are not in conflict.

\begin{figure}[t]
    \centering
\includegraphics[width=\linewidth]{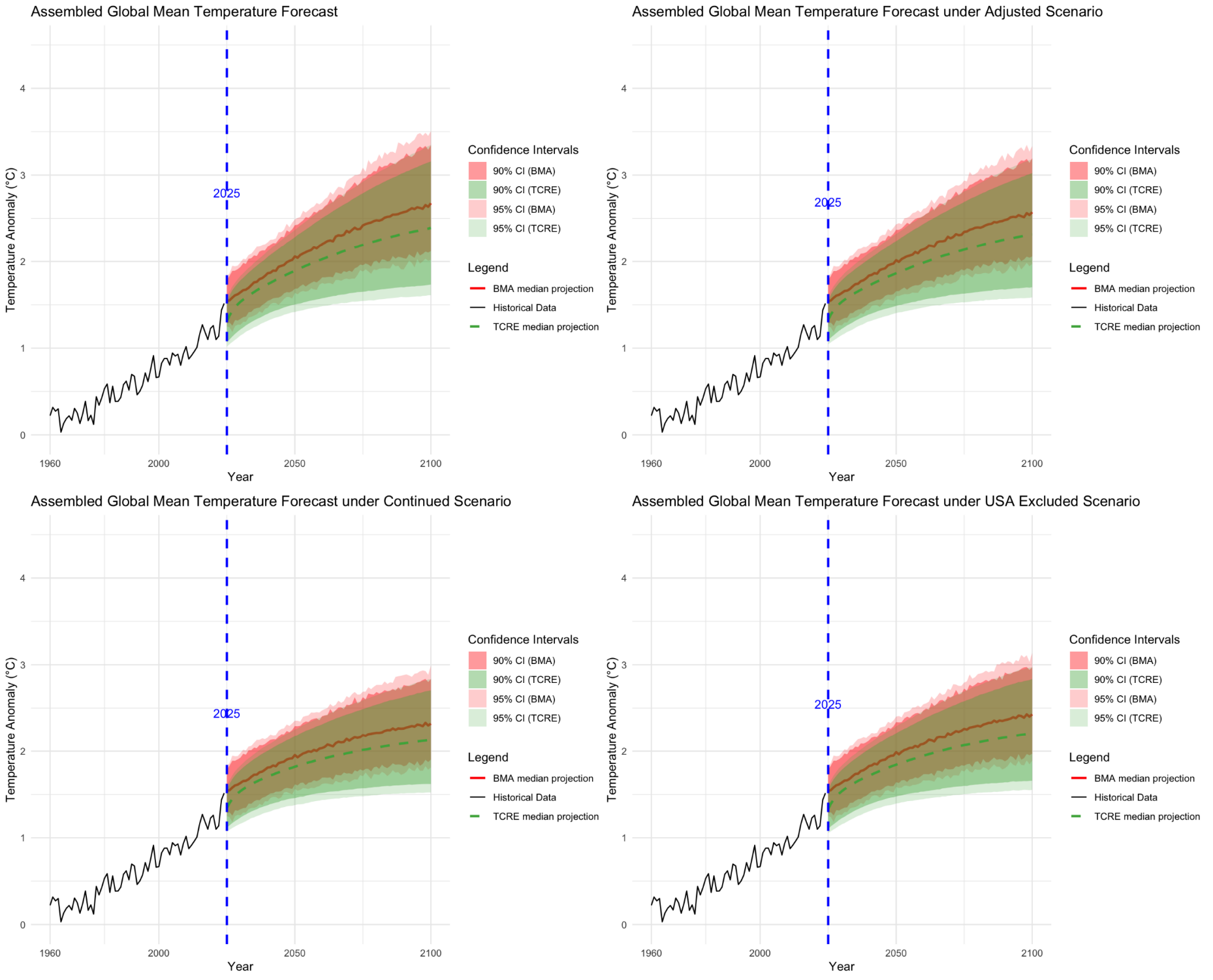}
    \caption{Temperature anomaly projections from 2025 to 2100 under four emission scenarios using the BMA framework: \textit{None} (top left) assumes a continuation of current trends, i.e., Figure \ref{projection_none}; \textit{Adjusted} (top right) assumes countries meet their Paris Agreement targets through 2030 only; \textit{Continued} (bottom left) assumes sustained compliance beyond 2030; and \textit{USA Excluded} (bottom right) assumes the United States withdraws from the agreement.}
    \label{fig:4 scenario}
\end{figure}

In addition to these baseline scenarios, which derive from current \(\mathrm{CO_2}\) emission trends, we further evaluated three alternative trajectories reflecting varying degrees of adherence to international climate agreements \citep{liu2021country}. We examined scenarios in which (i)~countries follow their Paris Agreement commitments up to 2030, (ii)~such commitments persist through 2100, and (iii)~a pathway in which the United States withdraws from the agreement. The temperature anomaly projection from 2025 to 2100 under four different scenarios are shown in Figure \ref{fig:4 scenario}.  Although each scenario presents different absolute levels of projected warming, all four scenarios exhibit similar trends under the BMA framework.

\subsection{Variance decomposition}

Since the uncertainty in BMA-based temperature change projections arises from two primary sources—(i) uncertainty in projected \(\mathrm{CO_2}\) emissions and (ii) uncertainty in the estimated parameters—we conducted a variance decomposition analysis to quantify their respective contributions.

The original projection results give a predictive standard deviation of global average temperature change to 2100 of 0.37\degree C, corresponding to a  variance of $0.37^2 = 0.14$. To isolate the effect of \(\mathrm{CO_2}\) emissions uncertainty, we first fixed a single emissions trajectory—the median trajectory among the 1,000 projected pathways—while continuing to sample model parameters from the posterior distribution. Under this setting, the variance was reduced to 0.09, indicating that \(\mathrm{CO_2}\) emissions variability contributes about 36\% of the total variance. Next, we further constrained the model by fixing both the \(\mathrm{CO_2}\) trajectory and the parameters to their posterior mean values, leaving only the residual uncertainty from the AR(1) autocorrelated error term. This final step resulted in a variance of 0.03, indicating that the error term accounts for about 21\% of the total variance. 

We thus find that, among the sources of uncertainty, the biggest single source is parameter uncertainty—primarily driven by climate sensitivity—which contributes 43\% of the total variance. This underscores the importance of improving climate sensitivity estimation to enhance predictive accuracy for probabilistic climate modeling.



\section{Discussion}
Our study presents a Bayesian Model Averaging (BMA) approach for weighting CMIP climate models based on their consistency with observational data.  While the weighting mechanism is inherent to BMA, the resulting weights provide a data-driven metric to quantify the agreement between individual models and historical records. This approach preserves ensemble diversity while anchoring projections in observational evidence. In practice, this is implemented through simulated temperature trajectories, which allow us to combine information across different models and produce an overall prediction with uncertainty.


Previous studies have explored different model weighting techniques. \cite{massoud2023bayesian} applied BMA to CMIP6 models to optimize alignment with IPCC's ECS and TCR ranges, but their approach relied on predefined climate sensitivity constraints rather than direct validation against observational data. In contrast, our study directly incorporates real-world observations to guide model selection, making it a more directly data-driven method. Similarly, \cite{wootten2023assessing} compared their version of BMA with skill-based weighting methods and found that it was more effective at reducing uncertainty, though it was sensitive to input assumptions. Our results confirm this sensitivity and also show that BMA-based projections are higher than unweighted CMIP6 ensembles. Our BMA approach is also one way to address  concerns about ``hot model biases'' in CMIP6 \citep{boyles2024approaches}, in that models with warm biases relative to observational data are automatically downweighted by the BMA method.

One advantage of our BMA approach is that it allows for a fully probabilistic estimation of climate sensitivity while maintaining internal model consistency. Methods that rely on hard thresholding, such as ECS- and TCR-based filtering, can discard valuable model information. By contrast, BMA adjusts model weights continuously based on statistical likelihood, preventing abrupt model exclusions that could degrade results. 

However, our approach is not without limitations. First, BMA is relatively computationally expensive, requiring Markov Chain Monte Carlo (MCMC) sampling to estimate posterior distributions. This could pose challenges for real-time climate assessments, although the computation for BMA is small compared to that required to run the CMIP models themselves. Second, our method assumes that observational records provide an unbiased reference for evaluating model performance. If historical observations contain systematic biases (e.g., due to incomplete coverage of ocean temperatures), the BMA weighting scheme may inherit these biases. 

A potential direction for future research is to explore alternative climate variables beyond global mean temperature, such as regional precipitation or extreme weather events. This could further enhance the applicability of BMA for climate risk assessment.

\begin{acks}[Acknowledgments]
The authors would like to thank Kyle Armour and David Battisti for helpful conversations. Raftery's research was partially supported by the Blumstein-Jordan Professorship at the University of Washington. ChatGPT was used to refine the wording of some paragraphs, but not for any substantive purposes.
\end{acks}

\begin{supplement}
\stitle{Supplementary material}
\sdescription{Formal derivations for the predictive distribution of future temperature and temperature change under an AR(1) error structure, the corresponding expectations and variances, and the Monte Carlo procedure are provided in the supplementary material.}
\end{supplement}

\begin{supplement}
\stitle{Data and code availability}
\sdescription{Software, code and data to replicate this analysis are available at the GitHub repository: \url{https://github.com/jitongj/BMA_CMIP6}.}

\end{supplement}

\bibliographystyle{imsart-nameyear}  
\bibliography{sample} 
\newpage

\end{document}


\maketitle

\section{Distribution of temperature and
temperature change}

After obtaining the posterior joint distribution $ \begin{bmatrix} \alpha \\ \beta \end{bmatrix} \mid \mathcal{D, \tilde D} \sim \mathcal{N} \left( \begin{bmatrix} \hat{\alpha} \\ \hat{\beta} \end{bmatrix}, \Sigma \right)$
, we apply it to find the predictive distribution of future temperature change. Our basic model is:
\begin{align*}
y_t &= \alpha + \beta x_t + u_{t}, \\
u_{t} &= \phi u_{t-1} + \varepsilon_{t},
\end{align*}
where
\begin{equation*}
\varepsilon_{t} \overset{iid}{\sim} N(0, \sigma^2).
\end{equation*}

The transformed model is:
\begin{equation*}
y'_t = \alpha' + \beta' x'_t + \varepsilon_{t},
\end{equation*}
where

\begin{align*}
    y'_t &= y_t - \phi y_{t-1},\\
    x'_t &= x_t - \phi x_{t-1},\\
    \alpha' &= \alpha (1 - \phi),\\ 
    \beta' &= \beta.
\end{align*}

Let the temperature and cumulative CO2 emission at year $t$ be $y_{t}$ and $x_{t}$ respectively, 
To calculate the temperature change, $\Delta T$, from year $t_1$ to $t_2$ using a basic model, we derive the following expression:
\begin{align*}
    \Delta T &= y_{t_2} - y_{t_1} \\
    &= \beta \cdot (x_{t_2} - x_{t_1}) + (u_{t_2} - u_{t_1}) \\
    &= \beta \Delta x + \Delta u
\end{align*}
where
\begin{align*}
    \Delta x &= (x_{t_2} - x_{t_1}), \\
    \Delta u &= (u_{t_2} - u_{t_1}).
\end{align*}

\subsection{Finding the distribution of $\Delta T$}
Assume $u_0 = 0$.

\begin{align*}
    u_1 &= \varepsilon_1 \\
    u_2 &= \phi u_1 + \varepsilon_2 = \phi \varepsilon_1 + \varepsilon_2 \\
    u_3 &= \phi u_2 + \varepsilon_3 = \phi (\phi \varepsilon_1 + \varepsilon_2) + \varepsilon_3 = \phi^2 \varepsilon_1 + \phi \varepsilon_2 + \varepsilon_3 \\
    &\vdots \\
    u_{t_1} &= \phi^{t_1-1} \varepsilon_1 + \phi^{t_1-2} \varepsilon_2 + \dots + \phi \varepsilon_{t_1-1} + \varepsilon_{t_1}
\end{align*}

Take $t_2 = t_1 + 1$.

\begin{align*}
    \Delta u &= u_{t_2} - u_{t_1} \\
    &= \phi u_{t_1} + \varepsilon_{t_1+1} - u_{t_1} \\
    &= (\phi - 1) u_{t_1} + \varepsilon_{t_1+1} \\
    &= (\phi - 1) \left[ \phi^{t_1-1} \varepsilon_1 + \phi^{t_1-2} \varepsilon_2 + \dots + \phi \varepsilon_{t_1-1} + \varepsilon_{t_1} \right] + \varepsilon_{t_1+1}
\end{align*}

Since $\varepsilon_t \overset{iid}{\sim} N(0, \sigma^2)$,

\begin{align*}
    \Delta u| \phi,\sigma^2 &\sim N(0, \sigma^2 \left[ (\phi - 1)^2 \left( \phi^{2t_1-2} + \phi^{2t_1-4} + \dots + \phi^2 + 1 \right) + 1 \right]) \\
    \Delta u| \phi,\sigma^2 &\sim N(0, \sigma^2 \left[ (\phi - 1)^2 \cdot \frac{1 - \phi^{2t_1}}{1 - \phi^2} + 1 \right]) \quad \text{(} \phi \neq 1) \\
\end{align*}

Now, take $t_2 = t_1 + m$.

\begin{align*}
    \Delta u &= u_{t_2} - u_{t_1} \\
    &= \phi^m u_{t_1} + \phi^{m-1} \varepsilon_{t_1+1} + \dots + \phi \varepsilon_{t_1+m-1} + \varepsilon_{t_1+m} - u_{t_1} \\
    &= (\phi^m - 1) \left[ \phi^{t_1-1} \varepsilon_1 + \phi^{t_1-2} \varepsilon_2 + \dots + \phi \varepsilon_{t_1-1} + \varepsilon_{t_1} \right] \\
    &+ \phi^{m-1} \varepsilon_{t_1+1} + \phi^{m-2} \varepsilon_{t_1+2} + \dots + \varepsilon_{t_1+m}
\end{align*}

Therefore, 
\begin{align*}
    \Delta u| \phi,\sigma^2 &\sim N(0, \sigma^2 (\phi^m - 1)^2 \times \frac{1 - \phi^{2t_1}}{1 - \phi^2} + \sigma^2 \times \frac{1 - \phi^{2m}}{1 - \phi^2})
\end{align*}
and we have 
\begin{align*}
    u_t| \phi,\sigma^2 &\sim N(0,  \sigma^2 \times \frac{1 - \phi^{2t}}{1 - \phi^2})
\end{align*}

Based on $\Delta T = \beta \Delta x + \Delta u$, we can get:

\begin{align*}
    \mathbb{E}[\Delta T] &= \mathbb{E}[\beta  \Delta x  + \Delta u] \\
    &= \mathbb{E}[\beta  \Delta x ] +\mathbb{E}[\Delta u]\\
    &= \Delta x \mathbb{E}[\beta] + \mathbb{E}_{\phi, \sigma^2}[ \mathbb{E}[\Delta u \mid \phi, \sigma^2] ] \\
    &= \Delta x \hat{\beta} + \mathbb{E}[0]\\
    &= \Delta x \hat{\beta}\\
\end{align*}

\begin{align*}
\text{Var} (\Delta T) &= \mathbb{E}_{\beta,\phi,\sigma^2}\left[\text{Var}(\Delta T \mid \beta, \phi, \sigma^2)\right] 
+ \text{Var}_{\beta,\phi,\sigma^2}(\mathbb{E}[\Delta T \mid \beta, \phi, \sigma^2]) \\
&= \mathbb{E}_{\beta,\phi,\sigma^2}\left[\text{Var} \left(\Delta x \beta + \Delta u \mid \beta, \phi, \sigma^2 \right)\right] 
+ \text{Var}_{\beta,\phi,\sigma^2} \left[\mathbb{E}\left(\Delta x \beta + \Delta u \mid \beta, \phi, \sigma^2 \right)\right] \\
&= \mathbb{E}_{\beta,\phi,\sigma^2}\left[\text{Var} \left(\Delta u \mid \beta, \phi, \sigma^2\right)\right]
+ \text{Var}_{\beta,\phi,\sigma^2} \left(\Delta x \beta + \mathbb{E}\left[\Delta u \mid \beta, \phi, \sigma^2\right]\right) \\
&= \mathbb{E}_{\phi,\sigma^2}\left[\text{Var} \left(\Delta u \mid \phi, \sigma^2\right)\right]
+ \text{Var}_{\beta,\phi,\sigma^2} \left(\Delta x \beta + \mathbb{E}\left[\Delta u \mid \phi, \sigma^2\right]\right) \\
&= \mathbb{E}_{\phi,\sigma^2} \left[\sigma^2 (\phi^m - 1)^2 \times \frac{1 - \phi^{2t_1}}{1 - \phi^2} + \sigma^2 \times \frac{1 - \phi^{2m}}{1 - \phi^2}\right]
+ \text{Var}_{\beta} \left(\Delta x \beta\right) \\
&= \mathbb{E}_{\phi,\sigma^2} \left[\sigma^2 (\phi^m - 1)^2 \times \frac{1 - \phi^{2t_1}}{1 - \phi^2} + \sigma^2 \times \frac{1 - \phi^{2m}}{1 - \phi^2}\right]
+\Delta x ^2 \Sigma_{22}
\end{align*}

\subsection{Finding the distribution of $y_t$}

Given $y_t = \alpha + \beta x_t + u_{t}$ and $u_t| \phi,\sigma^2 \sim N(0,  \sigma^2 \times \frac{1 - \phi^{2t}}{1 - \phi^2})$:

\begin{align*}
\mathbb{E}[y_t] &= \mathbb{E}[\alpha + \beta x_t + u_t] \\
                &= \mathbb{E}[\alpha] + \mathbb{E}[\beta] x_t + \mathbb{E}[u_t] \\
                &= \mathbb{E}[\alpha] + \mathbb{E}[\beta] x_t + \mathbb{E}_{\phi, \sigma^2}[\mathbb{E}[u_t \mid \phi, \sigma^2]]\\
                &= \mathbb{E}[\alpha] + \mathbb{E}[\beta] x_t + \mathbb{E}_{\phi, \sigma^2}[\mathbb{E}[u_t \mid \phi, \sigma^2]]\\
                &= \hat{\alpha} + \hat{\beta} x_t
\end{align*}

Since $u_t| \phi,\sigma^2 \sim N(0,  \sigma^2 \times \frac{1 - \phi^{2t}}{1 - \phi^2})$, we can derive the variance based on Law of Total Variance as following:

\begin{align*}
\text{Var}(y_t) &= \text{Var}(\alpha + \beta x_t + u_t) \\
&= \mathbb{E}_{\alpha,\beta,\phi,\sigma^2} \left[ \text{Var}(\alpha + \beta x_t + u_t \mid \alpha, \beta, \phi, \sigma^2) \right] \\
&\quad + \text{Var}_{\alpha,\beta,\phi,\sigma^2} \left( \mathbb{E}[\alpha + \beta x_t + u_t \mid \alpha, \beta, \phi, \sigma^2] \right) \\
&= \mathbb{E}_{\alpha, \beta, \phi,\sigma^2} \left[ \text{Var}(u_t \mid \phi, \sigma^2) \right] 
+ \text{Var}_{\alpha,\beta,\phi,\sigma^2} \left( \alpha + \beta x_t + \mathbb{E}[u_t \mid \phi, \sigma^2] \right) \\
&= \mathbb{E}_{\phi,\sigma^2} \left[ \sigma^2 \times \frac{1 - \phi^{2t}}{1 - \phi^2} \right] 
+ \text{Var}_{\alpha,\beta} \left( \alpha + \beta x_t + 0\right) \\
&= \mathbb{E}_{\phi,\sigma^2} \left[ \sigma^2 \times \frac{1 - \phi^{2t}}{1 - \phi^2} \right] 
+ \text{Var}_{\alpha,\beta} \left( \alpha + \beta x_t \right) \\
&= \mathbb{E}_{\phi,\sigma^2} \left[ \sigma^2 \times \frac{1 - \phi^{2t}}{1 - \phi^2} \right] 
+ \text{Var}(\alpha) + \text{Var}(\beta x_t) + 2 \text{Cov}(\alpha, \beta x_t) \\
&= \mathbb{E}_{\phi,\sigma^2} \left[ \sigma^2 \times \frac{1 - \phi^{2t}}{1 - \phi^2} \right] 
+ \Sigma_{1,1} + x_t^2 \Sigma_{2,2} + 2 x_t \Sigma_{1,2}
\end{align*}

We can approximate $y_t$ by a normal distribution with mean equals to $\hat{\alpha} + \hat{\beta} x_t$ and variance equal to $\mathbb{E}_{\phi,\sigma^2} \left[ \sigma^2 \times \frac{1 - \phi^{2t}}{1 - \phi^2} \right] 
+ \Sigma_{1,1} + x_t^2 \Sigma_{2,2} + 2 x_t \Sigma_{1,2}$.

\subsection{Calculating $\mathbb{E}_{\phi,\sigma^2} 
\left[\sigma^2 (\phi^m - 1)^2 \times \frac{1 - \phi^{2t_1}}{1 - \phi^2} + \sigma^2 \times \frac{1 - \phi^{2m}}{1 - \phi^2}\right]$ and $\mathbb{E}_{\phi,\sigma^2} \left[ \sigma^2 \times \frac{1 - \phi^{2t}}{1 - \phi^2} \right]$}

To calculate $\mathbb{E}_{\phi,\sigma^2} \left[\sigma^2 (\phi^m - 1)^2 \times \frac{1 - \phi^{2t_1}}{1 - \phi^2} + \sigma^2 \times \frac{1 - \phi^{2m}}{1 - \phi^2}\right]$ in $\text{Var} (\Delta T)$, and $\mathbb{E}_{\phi,\sigma^2} \left[ \sigma^2 \times \frac{1 - \phi^{2t}}{1 - \phi^2} \right]$ in $\text{Var} (y_t)$, we sample $\alpha$ and $\beta$ from $ \mathcal{N} \left( \begin{bmatrix} \hat{\alpha} \\ \hat{\beta} \end{bmatrix}, \Sigma \right).$\\

For every sampled $(\alpha_j, \beta_j)$, we can get their corresponding $\phi_j$, which yields $\sigma_j$ given $u_{t} = \phi u_{t-1} + \varepsilon_{t}$ where $\varepsilon_{t} \overset{iid}{\sim} N(0, \sigma^2).$\\

Sampling 1,000 times, i.e. $j = 1 \dots 1000$, $\mathbb{E}_{\phi,\sigma^2} \left[\sigma^2 (\phi^m - 1)^2 \times \frac{1 - \phi^{2t_1}}{1 - \phi^2} + \sigma^2 \times \frac{1 - \phi^{2m}}{1 - \phi^2}\right]$ is equal to:

\begin{align*}
    &=\frac{\sum_{j=1}^{1000} \sigma_j^2 (\phi_j^m - 1)^2 \times \frac{1 - \phi_j^{2t_1}}{1 - \phi_j^2} + \sigma_j^2 \times \frac{1 - \phi_j^{2m}}{1 - \phi_j^2}}{1000}
\end{align*}

and 
\begin{equation*}
    \mathbb{E}_{\phi,\sigma^2} \left[ \sigma^2 \times \frac{1 - \phi^{2t}}{1 - \phi^2} \right] =\frac{\sum_{j=1}^{1000} \sigma_j^2 \times \frac{1 - \phi_j^{2t}}{1 - \phi_j^2}}{1000} .
\end{equation*}

\section{Results for model selection (2) all-realization aggregation}
\begin{figure}
    \centering
    \includegraphics[width=0.8\linewidth]{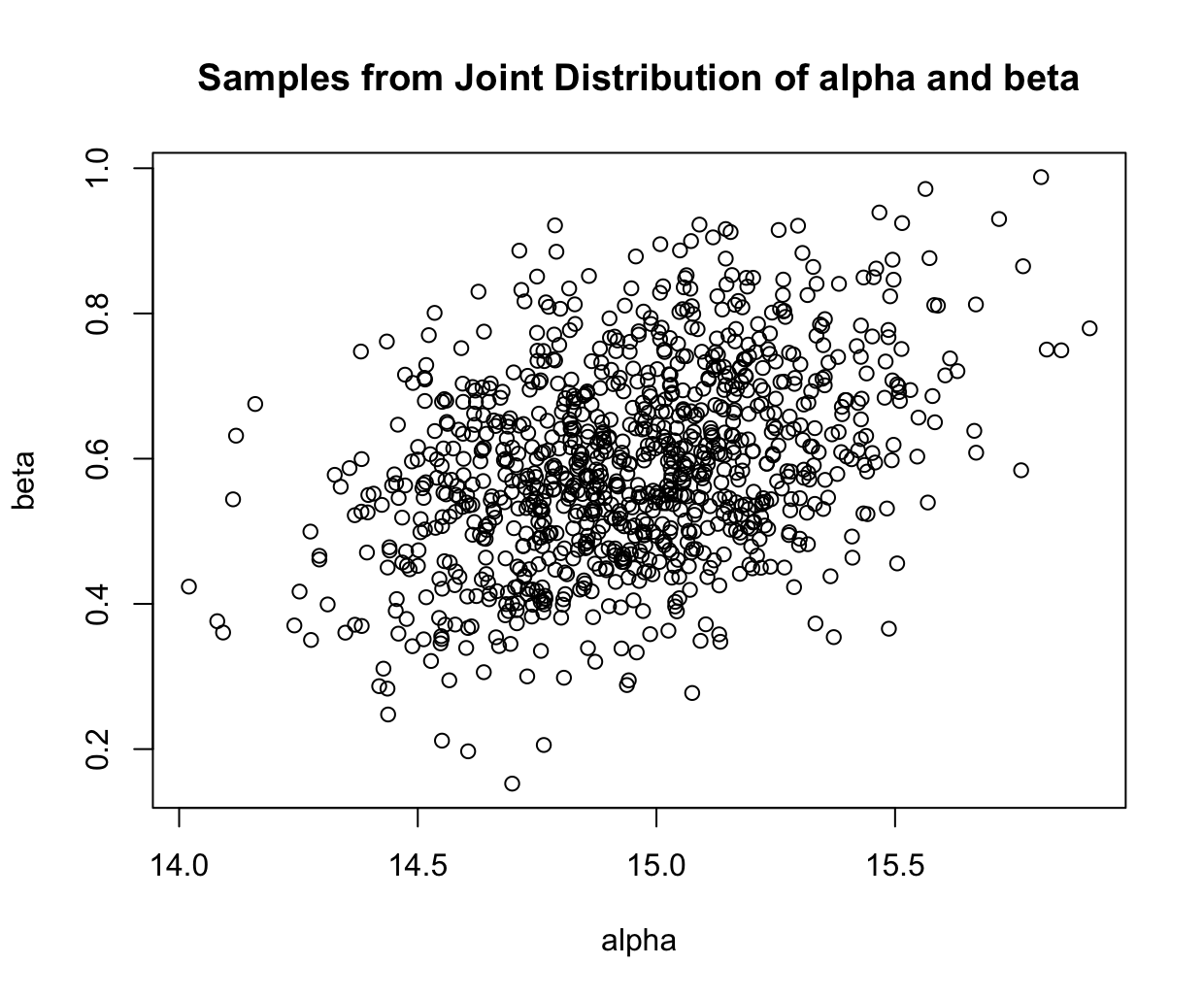}
    \caption{Samples from the joint posterior distribution of $\alpha$ and $\beta$ obtained from the BMA model based on model selection (2). The scatter plot reveals a positive dependence between the two parameters, consistent with the estimated posterior correlation of 0.49.}
    \label{fig:alpha_beta}
\end{figure}
\subsection{Model averaging}
Applying BMA to all 51 realizations under the aggregation-based procedure described in the main text yields posterior mean estimates of $\alpha = 14.95$ and $\beta = 0.59$~$^\circ$C per 1000 Gt $\mathrm{CO_2}$. The corresponding posterior standard deviations are 0.30 for $\alpha$ and 0.13 for $\beta$, with a posterior correlation of 0.40 (see Figure~\ref{fig:alpha_beta}). Relative to the TCRE benchmark, the estimate of $\beta$ shows a 31\% increase in the posterior mean and a 28\% reduction in posterior standard deviation. These results are  consistent with those obtained under the random realization sampling approach reported in the main text.

\begin{figure}[t]
    \centering
    \includegraphics[width=0.8\linewidth]{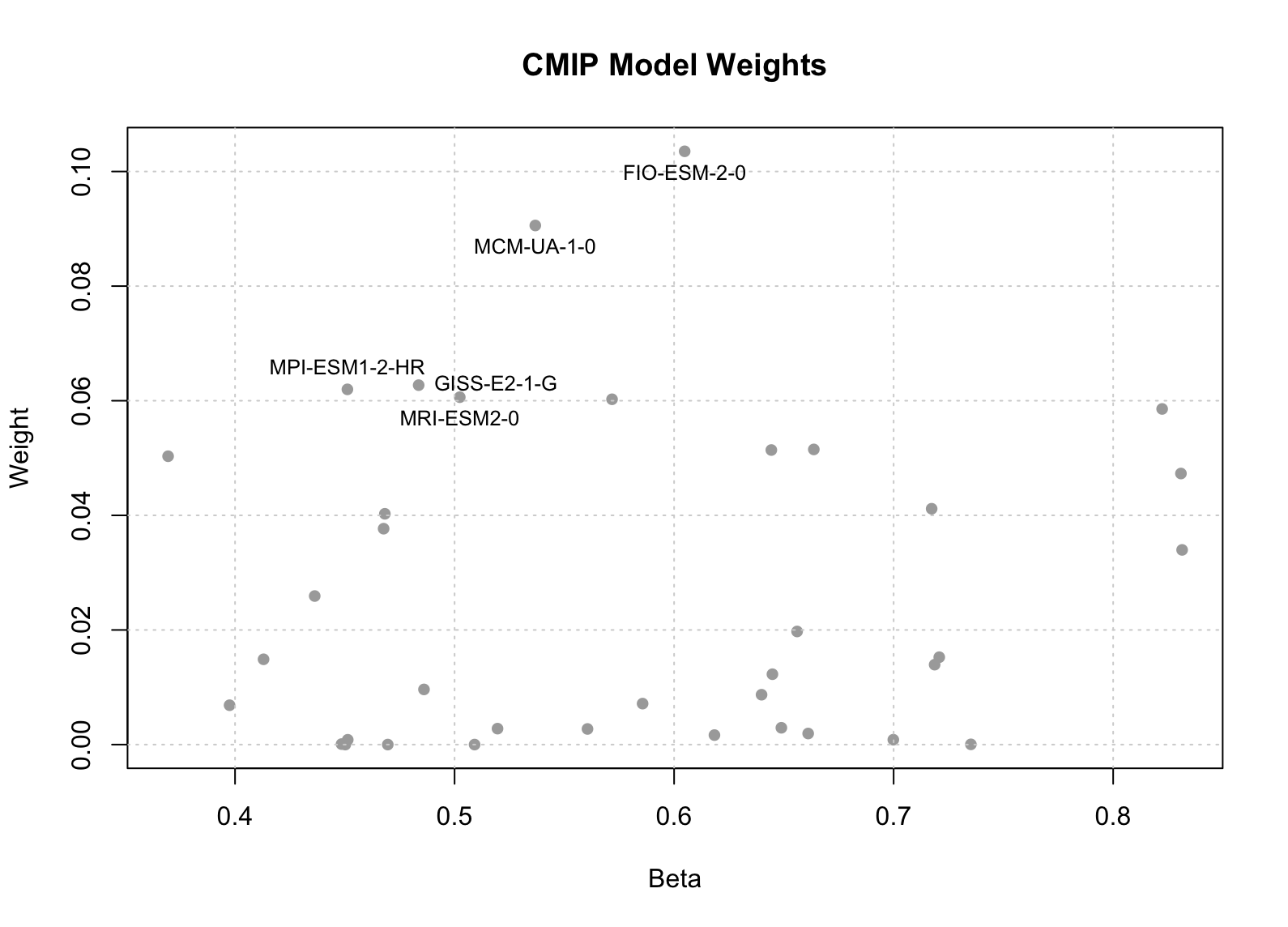}
    \caption{Posterior model weights from the BMA procedure applied to 37 CMIP6 climate models, where each weight reflects the model’s relative fit to observed historical temperatures. The regression coefficients $\beta_k$ (i.e., estimates of climate sensitivity) are estimated via OLS for each model.} 
    \label{weights}
\end{figure}

Figure~\ref{weights} displays the resulting BMA weights aggregated at the model-family level. The weights exhibit substantial heterogeneity across models. The model family ``FIO-ESM-2-0'' receives the largest weight 0.10, followed by ``MCM-UA-1-0'' (0.09), while ``GISS-E2-1-G'', ``MPI-ESM1-1-HR'', and ``MRI-ESM2-0'' each receive weights of 0.06. The set of top-weighted models is consistent with that identified under the random sampling approach.

\begin{figure}
    \centering
    \includegraphics[width=0.8\linewidth]{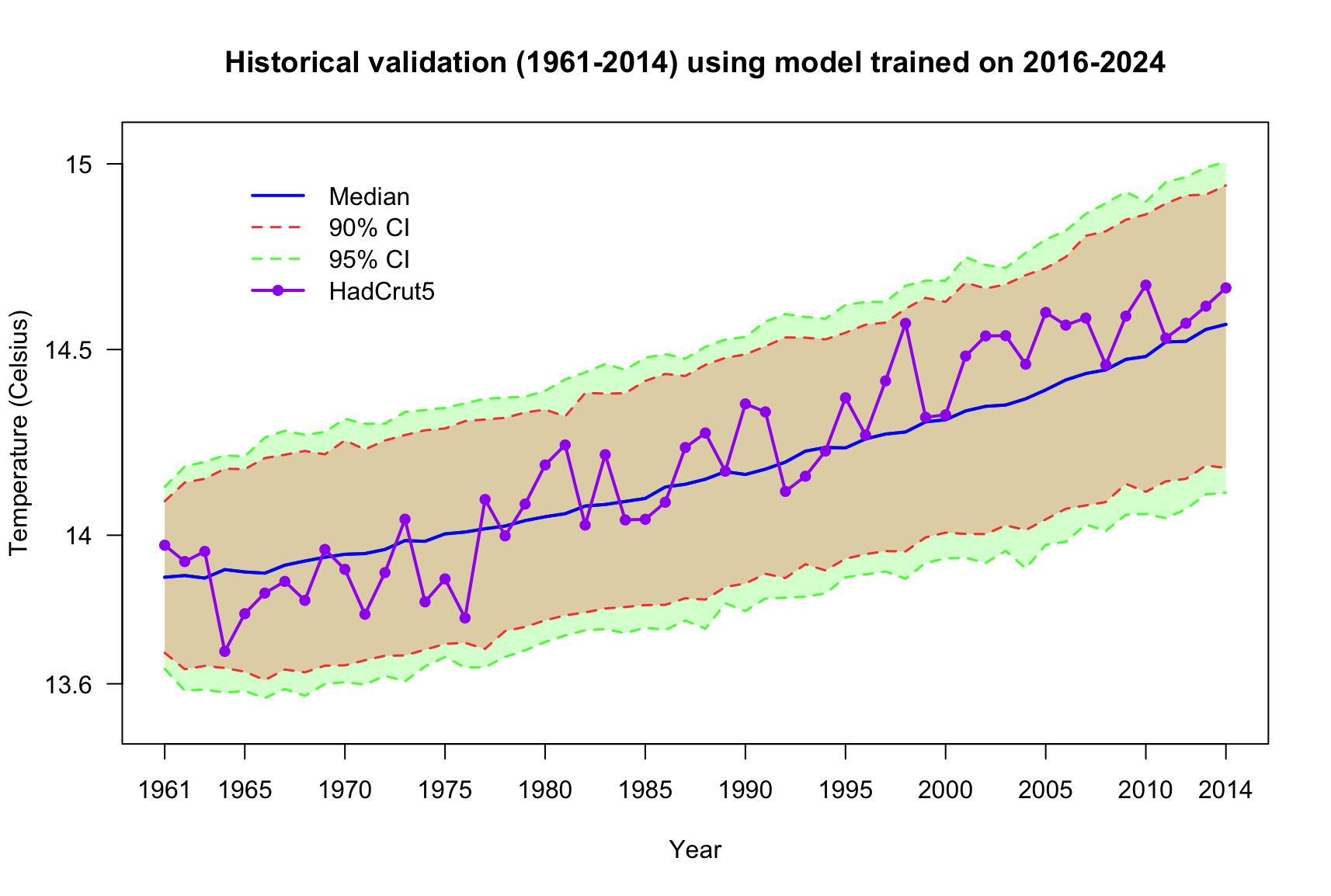}
    \caption{Historical validation (1961--2014) using the model trained on 2016--2024 data. Posterior median (blue), 90\% and 95\% credible intervals (red and green dashed lines), and observed temperatures from HadCRUT5 (purple). The model captures the long-term trend and provides well-calibrated uncertainty.}
    \label{fig:his_validation}
\end{figure}

\begin{figure}[t]
    \centering
\includegraphics[width=\linewidth]{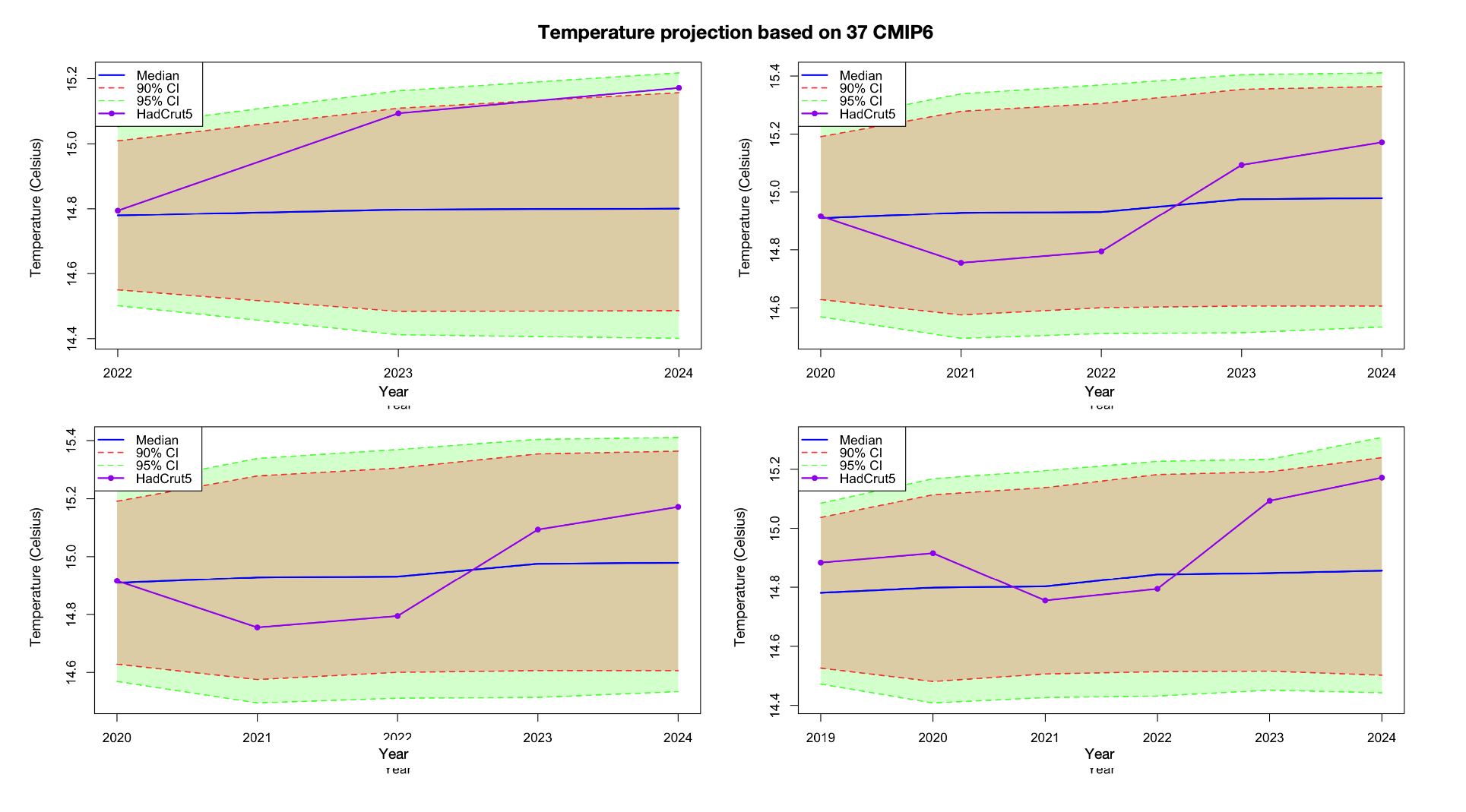}
    \caption{Out-of-sample validation of the BMA-based temperature projections using holdout periods ending in 2018, 2019, 2020, and 2021. The posterior median and 90\% and 95\% credible intervals are compared to observed global temperature anomalies from HadCRUT5. Results demonstrate good calibration and predictive accuracy.
}
    \label{fig:os_validation}
\end{figure}

\subsection{Validation, projection and variance decomposition}

Under the aggregation-based procedure, the validation results are consistent with those reported in the main text and are therefore not repeated here (See Figure \ref{fig:his_validation} and \ref{fig:os_validation}).

The corresponding BMA projections are also similar to those obtained under random realization sampling. The posterior median temperature increase by 2100 is $2.68^\circ\mathrm{C}$, with a 90\% credible interval of $(2.12, 3.36)$. The total predictive variance at the year 2100 is estimated to be $0.14$.  

Again, we decompose the predictive variance of the projected temperature change into contributions from (i) emissions uncertainty, (ii) parameter uncertainty, and (iii) residual variability induced by the AR(1) error structure. To assess the contribution of emissions uncertainty, we fix the $\mathrm{CO_2}$ trajectory at the median of the 1{,}000 simulated pathways and sample parameters from their posterior distribution. Under this setting, the variance is reduced to $0.10$, implying that emissions uncertainty accounts for approximately 29\% of the total variance. Next, we fix the model parameters at their posterior mean values, leaving only the AR(1) error term. The resulting variance is $0.03$, indicating that residual variability contributes approximately 21\% of the total variance. The remaining variation, approximately 50\%, is attributable to parameter uncertainty. This is primarily driven by uncertainty in climate sensitivity and represents the largest single component of the predictive variance.

Overall, these results are consistent with those obtained under the random realization sampling approach, indicating that the main findings are not sensitive to the model selection strategy.